\documentclass[aps,showpacs,preprint,preprintnumber,nofootinbib,amsmath,amssymb,ascmac,bm,12pt]{revtex4}
\usepackage{bm}
\usepackage[dvipdfm]{graphicx}
\begin{document}
\title{\vspace*{0.5cm}
Extremal Charged Black Holes with a Twisted Extra Dimension
}
\author{
${}^{1}$Takamitsu Tatsuoka\footnote{E-mail:tatsuoka@sci.osaka-cu.ac.jp},
${}^{1}$Hideki Ishihara\footnote{E-mail:ishihara@sci.osaka-cu.ac.jp},
${}^{2}$Masashi Kimura\footnote{E-mail:mkimura@yukawa.kyoto-u.ac.jp}
and
${}^{1}$Ken Matsuno\footnote{E-mail:matsuno@sci.osaka-cu.ac.jp}
}
\affiliation{
${}^{1}$Department of Mathematics and Physics, Osaka City University, Sumiyoshi, Osaka 558-8585, Japan
\\
${}^{2}$Yukawa Institute for Theoretical Physics, Kyoto University, Kyoto 606-8502, Japan
}
\begin{abstract}
We construct odd-dimensional extremal charged black hole solutions with a twisted S$^1$ as an extra dimension on generalized Euclidean Taub-NUT spaces. There exists a null hypersurface where an expansion for an outgoing null geodesic congruence vanishes, then these spacetimes look like black holes. We show that the metrics admit C$^0$ extension across the horizon, but some components of Riemann curvature diverge there if the dimension is higher than five. The singularity is relatively mild so that an observer along a free-fall geodesic can traverse the horizon. We also show solutions with a positive cosmological constant. 
\end{abstract}

\preprint{OCU-PHYS 348}
\preprint{AP-GR 91}
\preprint{YITP-11-42}


\date{\today}
\maketitle

\section{Introduction}

In recent years, studies on higher dimensional black holes have attracted 
much attention. 
After the discovery of black ring solutions~\cite{Emparan:2001wn}, 
it has been shown that higher-dimensional black objects admit 
various horizon topologies in contrast to the four-dimensional case. 
In addition to black holes with spherical horizons~\cite{Tangherlini:1963bw,Myers:1986un}, 
there are many exotic solutions of the Einstein equation in higher dimensions, 
e.g., black strings, rings, Saturns, and
di-rings~\cite{Emparan:2001wn,Mishima:2005id,Figueras:2005zp,Pomeransky:2006bd,
Elvang:2007rd,Iguchi:2007is,Izumi:2007qx,Elvang:2007hs,Evslin:2008py}.

From a realistic point of view, the extra dimensions need to be 
compactified to reconcile the higher-dimensional theory of gravity with 
our apparently four-dimensional world. 
Then, it is important to consider higher-dimensional black holes with compact 
extra dimensions, i.e., Kaluza-Klein black holes. 
In five dimensions, for example, exact Kaluza-Klein black hole solutions 
with a twisted extra dimension, 
i.e., squashed Kaluza-Klein (SqKK) black holes, 
are constructed~\cite{Dobiasch:1981vh, Gibbons:1985ac, 
Gauntlett:2002nw,Gaiotto:2005gf,
Ishihara:2005dp, 
Yazadjiev:2006iv,
Nakagawa:2008rm, Tomizawa:2008hw, Tomizawa:2008rh, Stelea:2008tt, 
Tomizawa:2008qr,Bena:2009ev,Tomizawa:2010xq,Mizoguchi:2011zj}
in a class of cohomogeneity-one symmetry. 
These black holes have spherical horizon topologies, and they look like 
dimensionally reduced black holes in their asymptotic regions. 

On the basis of a Ricci-flat space with the Euclidean signature, 
extremal charged black hole solutions can be constructed 
by using harmonic functions on the base space~\cite{Myers:1986rx}.
If we take a hyperk\"ahler space as the base space, we obtain a supersymmetric 
black hole~\cite{Gauntlett:2002nw}. 
Many supersymmetric black object solutions are obtained explicitly, 
for example, 
black holes \cite{Breckenridge:1996is,Herdeiro:2002ft,Herdeiro:2003un, 
Brecher:2003wq,Ortin:2004af, 
Kim:2010bf}, 
black rings \cite{Elvang:2004rt,Elvang:2004ds,Elvang:2005sa,Bena:2005ni, 
Gaiotto:2005xt, Tomizawa:2007he,Camps:2008hb}, 
and 
multi-black objects \cite{Gauntlett:2004wh,Gauntlett:2004qy,
Ishihara:2006pb, Tomizawa:2008tj}.

In five dimensions, the extremal charged Kaluza-Klein 
black holes are constructed 
in such a way as to superpose harmonic functions 
on the flat Euclid space periodically, 
which are special cases of Majumdar-Papapetrou multi-black hole 
solutions, and to take an identification 
by the period~\cite{Myers:1986rx}.
The solutions have a  direct product structure of the extra S$^1$ 
with the base spacetime\footnote{
Generalizations to rotating black hole cases are 
shown in \cite{Maeda:2006hd}. 
}. 
The extremal charged Kaluza-Klein black hole solutions with a twisted S$^1$, 
which are extremal limit of SqKK black holes, 
are also constructed 
by using the Euclidean Taub-NUT space~\cite{Ishihara:2006iv, 
Gauntlett:2002nw,Gaiotto:2005gf,
Matsuno:2008fn,
Bena:2009ev}\footnote{
The solutions can be generalized to the cases with a positive cosmological constant 
\cite{London:1995ib,Klemm:2000vn,Behrndt:2004pn,Gutowski:2004ez, 
Gutowski:2004yv,Ishihara:2006ig,Ida:2007vi,Matsuno:2007ts,Kimura:2009er}.
}. 
The Taub-NUT space, which is a four-dimensional Ricci flat space, 
has the structure of S$^1$ fiber over the 
three-dimensional asymptotically flat base space, 
where the size of S$^1$ is finite at the infinity. 
Generalizations of the Taub-NUT space to higher-even-dimensional spaces are 
discussed in
\cite{Bais:1984xb,Awad:2000gg,Cvetic:2001zb,Clarkson:2002uj, 
Mann:2003zh,Lu:2004ya,Mann:2005gk,Awad:2005ff,Dehghani:2005zm,Dehghani:2006aa, 
Hendi:2008wq, Hatsuda:2009vj, Mann:2005ra}. 
In this paper,  
we focus on the generalized Taub-NUT spaces in higher dimensions, 
which are not hyperk\"ahler in general, as base spaces, 
and construct extremal charged solutions 
that have a twisted S$^1$ as an extra dimension explicitly. 
The solutions are extensions of  
the extremal charged SqKK black hole solutions in five dimensions to 
higher than five dimensions
\footnote{There are some attempts to obtain 
vacuum solutions of 
SqKK black holes in higher dimensions numerically~\cite{Lipert:2009sr}.
}.

It was pointed out that the Majumdar-Papapetrou multi-black hole solutions 
are analytic at the horizon in four dimensions, 
but not smooth in five dimensions, 
and have curvature singularities at the horizon 
in higher than five dimensions~\cite{Gibbons:1994vm,Welch:1995dh, 
Candlish:2007fh,Candlish:2009vy}. 
The smoothness of the horizon depends on the spacetime dimension. 
How about the smoothness of the horizons of Kaluza-Klein black holes? 
In five dimensions, the extremal charged 
SqKK black hole 
solutions
with a twisted S$^1$
have smooth horizons\footnote{
It was shown that five-dimensional extremal charged Kaluza-Klein multi-black 
hole solutions with a twisted S$^1$ are smooth at the horizons~\cite{Kimura:2008cq}.
}.
In this paper we examine whether the smoothness of the horizon is preserved in higher 
than five dimensions. 

This paper is organized as follows. 
In Sec.II, we construct extremal charged solutions with a twisted extra 
S$^1$ in odd dimensions explicitly. 
Sec.III and IV are devoted to an investigation of 
asymptotic structures of the solutions and 
the regularity at the horizon, respectively.
In Sec.V we present the solutions with a positive cosmological constant, 
and we conclude our works with discussions in Sec.VI. 

\section{Construction of extremal charged solutions to 
the Einstein-Maxwell equations}
\label{sec:constructingsolutions}

We consider the $D$-dimensional action of the 
Einstein-Maxwell theory
\begin{eqnarray}
S = \frac{1}{16\pi} \int d^D x \sqrt{-g} \left( R - F_{\mu\nu}F^{\mu\nu} \right),
\label{eq:Daction}
\end{eqnarray}
where $R$ is the Ricci tensor and $F_{\mu\nu} := \partial_\mu A_\nu - \partial_\nu A_\mu$
is the electromagnetic field tensor. The $D$-dimensional gravitational constant $G_D$ is set equal to one. 
Greek indices run over the $D$ time-space values 0, 1, $\cdots$, $D-1$. 
The variation of the action leads to the Einstein and Maxwell equations
\begin{eqnarray}
R_{\mu\nu} - \frac{1}{2}Rg_{\mu\nu} &=& 2 \left( F_{\mu\alpha}F_{\nu}^{\ \alpha} - \frac{1}{4}F_{\alpha\beta}F^{\alpha\beta}g_{\mu\nu} \right),  
\label{Einstein_eq}
\\
\nabla^{\nu}F_{\mu\nu} &=& 0.
\label{Maxwell_eq}
\end{eqnarray}
For $(D-1)$-dimensional Ricci-flat metrics $h_{ij}(x^k)$ 
(Latin indices run over the $(D-1)$ space values 1, $\cdots$, $D-1$), 
we consider $D$-dimensional metrics and Maxwell fields in the form of
\begin{eqnarray}
ds^2 &=& -H(x^i)^{-2}dt^2+H(x^i)^{2/(D-3)}h_{ij}dx^{i}dx^{j}, \\
A_{\mu} dx^{\mu} &=& \pm \sqrt{\frac{D-2}{2(D-3)}}\frac{dt}{H(x^i)}.
\end{eqnarray}
It is known that both the Einstein and Maxwell equations, \eqref{Einstein_eq}
and \eqref{Maxwell_eq}, are reduced to the 
Laplace equations on the base space with the metric $h_{ij}$ \cite{Myers:1986rx}, 
\begin{eqnarray}
\triangle_{h_{ij}}H(x^i)=0.
\label{eq:Laplace}
\end{eqnarray}

On the line of this fact, we construct Kaluza-Klein black hole 
solutions with a twisted extra dimension. 
In the five-dimensional case, such solutions can be constructed on the Euclidean 
Taub-NUT space 
\cite{
Ishihara:2006iv, Gauntlett:2002nw,Gaiotto:2005gf,Matsuno:2008fn,Bena:2009ev}. 
First, we generalize the Taub-NUT space to higher even dimensions than four, 
and construct the solutions by using a harmonic function on the space.

We consider $2(n+1)$-dimensional spaces with the metrics~\cite{Dehghani:2005zm}
\begin{eqnarray}
h_{ij}dx^i dx^j=\frac{dr^2}{F(r)}+r(r+2L)d\Sigma_{n}^{2}
+L^2 F(r)\left( d\chi +\omega_n\right) ^2,
\label{eq:D-1metric}
\end{eqnarray}
where $F$ is a function of $r$ and $L$ is a positive parameter which represents 
the size of the extra dimension, as will be shown later, 
and $d\Sigma_n^2$ is the metric on $\mathbb{C}P(n)$ 
defined recursively by
\begin{eqnarray}
d\Sigma_{I}^2 &=& (2I+2)\Biggl[d\xi_I^2 + \sin^2\xi_I \cos^2\xi_I 
\left( d\psi_I +\frac{1}{2I}\omega_{I-1}\right)^2+\frac{1}{2I}\sin^2\xi_I
d\Sigma_{I-1}^2\Biggr], \\
\omega_I &=& (2I+2)\sin^2\xi_I\left( d\psi_I +\frac{1}{2I}\omega_{I-1}\right), \\
d\Sigma_1^2 &=& 4\left( d\xi_1^2 +\sin^2\xi_1 \cos^2\xi_1 d\psi_1^2 \right), \\
\omega_1 &=& 4\sin^2\xi_1 d\psi_1,
\end{eqnarray}
where $0 \leq \xi_I \leq \pi/2, \,  0 \leq \psi_I \leq 2\pi, \, 0 \leq \chi \leq 4(n+1)\pi, (I=1,\cdots,n)$ are angular coordinates.
We require $h_{ij}$ to be Ricci-flat and regular at $r=0$.  
The Ricci flatness condition and the regularity condition are respectively given by
\begin{eqnarray}
&& [(2n-1)r^2+ 2(2n-1)Lr +  2nL^2] F + r (r+2L) [(r+L) \frac{dF}{dr} -1] = 0, \\
&& F(r) \to \frac{r}{(n+1)L} \quad \mbox{as} \quad r \to 0. 
\end{eqnarray}
Then, we find
\begin{eqnarray}
F(r)=\sum_{k=1}^{n}\frac{2^{n-k+1}n!}{(n+k)(n+k-1)(n-k)!(k-1)!}
\frac{L^{n-k}r^k}{(r+2L)^n}.
\label{eq:F}
\end{eqnarray}
We call, in this paper, the metrics (\ref{eq:D-1metric}) with 
\eqref{eq:F} the $2(n+1)$-dimensional generalized Taub-NUT spaces. 

Using generalized Taub-NUT spaces and harmonic functions $H$ on the spaces,
then we construct the $D=2(n+1)+1$ dimensional solutions on the base space with the metric $h_{ij}$ 
in the form
\begin{eqnarray}
&&ds^2 = - H(r)^{-2}dt^2
+H(r)^{1/n} \Bigl[ \frac{dr^2}{F(r)} +r(r+2L)d\Sigma_{n}^2
+L^2F(r)\left( d\chi +\omega_{n} \right)^2 
\Bigr],
\label{eq:Dmetric}\\
&&A_{\mu}dx^{\mu} = \pm \sqrt{\frac{2n+1}{4n}}\frac{dt}{H(r)},
\label{eq:Dfield}
\end{eqnarray}
where we assume that the function $H$ depends only on the radial coordinate $r$.
Then, the Laplace equations (\ref{eq:Laplace}) become
\begin{eqnarray}
r(r+2L)\frac{dF(r)}{dr}\frac{dH(r)}{dr}+F(r)\Biggl[ 2n(r+L)\frac{dH(r)}{dr}
+r(r+2L)\frac{d^2 H(r)}{dr^2}\Biggr] =0.
\end{eqnarray}
The solutions for a point source at $r=0$ are 
\begin{eqnarray}
H(r)=1+ \mu^{2n} \int_{r}^{\infty} dr \frac{1}{L[r(r+2L)]^{n}F(r)},
\label{eq:H}
\end{eqnarray}
where $\mu$ is an arbitrary constant, 
and we set another constant of integration suitably such that 
$H \rightarrow 1$ as $r\rightarrow \infty$. 
The Komar mass and charge for the solutions \eqref{eq:Dmetric} and \eqref{eq:Dfield} with \eqref{eq:F} and \eqref{eq:H} are 
\begin{eqnarray}
M &:=& - \frac{2n+1}{2n}\frac{1}{16\pi}
\int_{\infty}\nabla^{\mu}\xi^{\nu}dS_{\mu\nu} \nonumber \\
&=& \frac{2n+1}{2n}[2(n+1)]^{n+1}\frac{1}{8\pi}\mu^{2n}\Omega_{2n+1}, \\
Q &:=& - \frac{1}{8\pi} \int_{\infty} F^{\mu\nu} dS_{\mu\nu} \nonumber \\
&=& \pm \sqrt{\frac{2n+1}{n}}[2(n+1)]^{n+1}\frac{1}{8\pi}\mu^{2n}\Omega_{2n+1},
\end{eqnarray}
respectively, where $\Omega_{2n+1}$ denotes the area of the $(2n+1)$-dimensional unit sphere, 
and $\xi^{\mu} \partial_{\mu}$ is the timelike Killing vector field 
whose norm is $-1$ at $r=\infty$.
The solutions 
satisfy the extremality condition
\begin{eqnarray}
M = |Q| \sqrt{\frac{2n+1}{4n}}. 
\label{eq:extreme}
\end{eqnarray}
In the case $n=1$, the metrics coincide with the five-dimensional extremal 
Kaluza-Klein black hole metric on the Euclidean Taub-NUT space
~\cite{
Ishihara:2006iv, Gauntlett:2002nw,Gaiotto:2005gf,Matsuno:2008fn,Bena:2009ev}.

\section{Asymptotic structures}

\subsection{The $\mu=0$ case }
In the case of $\mu=0$, the metric \eqref{eq:Dmetric}
reduces to
\begin{eqnarray}
ds^2 = -dt^2
+ \frac{dr^2}{F(r)} +r(r+2L)d\Sigma_{n}^2
+L^2F(r)\left( d\chi +\omega_{n} \right)^2 .
\label{eq:Monopole}
\end{eqnarray}
If $r \ll L$, the metric (\ref{eq:Monopole}) behaves
\begin{eqnarray}
ds^2 \simeq -dt^2 + \frac{(n+1)L}{r}dr^2 + 2Lr d\Sigma_n^2 
+ \frac{Lr}{n+1}(d\chi+ \omega_n)^2.
\label{eq:Dmetricmu0r0}
\end{eqnarray}
Introducing the new radial coordinate $R$ defined by $r=R^2/(4(n+1)L)$, we find
\begin{eqnarray}
ds^2 \simeq -dt^2 + dR^2 + R^2 d\Omega_{2n+1}^2,
\label{eq:Dmetricmu0r0no2}
\end{eqnarray}
where
\begin{eqnarray}
d\Omega_{2n+1}^2 = \frac{1}{2(n+1)}d\Sigma_n^2 
+ \frac{1}{4(n+1)^2}\left( d\chi + \omega_n\right)^2
\end{eqnarray}
is the metric on the $(2n+1)$-dimensional unit sphere. 
Eq.(\ref{eq:Dmetricmu0r0no2}) shows that the spacetime (\ref{eq:Monopole}) 
has no conical singularity at $r=0$.

If $r \gg L$, the metrics (\ref{eq:Monopole}) behave as 
\begin{eqnarray}
ds^2_{D=5} &\simeq& -dt^2 + \left( 1+\frac{2L}{r} \right) dr^2 
+ r^2 \left( 1+\frac{2L}{r} \right) d\Sigma_1^2
+ L^2 \left( 1 -\frac{2L}{r} \right) \left( d\chi +\omega_1 \right)^2
\label{eq:5metricmu0rlarge} 
\end{eqnarray}
for $D=5$, and
\begin{eqnarray}
ds^2_{D>5} &\simeq& -dt^2 + 2(n-1)
\left( 1+\frac{2n}{2n-3}\frac{L^2}{r^2} \right) dr^2 
+ r^2 \left( 1 + \frac{2L}{r} \right) d\Sigma_n^2 \nonumber \\
&& + \frac{L^2}{2n-1} \left( 1-\frac{2n}{2n-3}\frac{L^2}{r^2} \right) 
\left( d\chi+ \omega_n\right)^2
\label{eq:Dmetricmu0rlarge}
\end{eqnarray}
for $D>5$. 
In both cases (\ref{eq:5metricmu0rlarge}) and (\ref{eq:Dmetricmu0rlarge}), 
the  angular parts of the metrics have the fiber bundle structures: an ${\rm S^1}$ fiber 
over the $\mathbb{C}P(n)$ base space. 
In the $n=1$ case, the base space is ${\rm S^2}$ and the metric is 
the Kaluza-Klein monopole solution~\cite{Gross:1983hb,Sorkin:1983ns}.
The spacetime with the metrics (\ref{eq:5metricmu0rlarge}) or (\ref{eq:Dmetricmu0rlarge}) 
commonly has a twisted ${\rm S^1}$ as a compactified extra dimension 
whose size is $4(n+1)\pi L/\sqrt{2n-1}$, respectively. 

\subsection{The $\mu\neq 0$ case }
If the parameter $\mu^{2n}$ is negative $\mu^{2n} < 0$, there exists a point $r=r_{s}>0$ where $H(r_s)=0$,
since the integral in the right hand side of (\ref{eq:H})
becomes arbitrary large as $r\to 0$.
In this case, the spacetime with the metric (\ref{eq:Dmetric}) has a timelike singularity 
at $r = r_s$ where the Kretschmann scalar $R_{\mu\nu\rho\sigma}R^{\mu\nu\rho\sigma}$ diverges. 
Then, here and henceforth, we consider the $\mu^{2n} >0$ case. 
We can assume $\mu>0$ without loss of generality. 

If $r \ll L$, the metric (\ref{eq:Dmetric}) behaves
\begin{eqnarray}
ds^2 &\simeq& 
- \Biggl[ 1 + \frac{n+1}{n}\frac{\mu^{2n}}{(2Lr)^n} \Biggr]^{-2}dt^2 
\nonumber \\
&& +\Biggl[ 1 + \frac{n+1}{n}\frac{\mu^{2n}}{(2Lr)^n} \Biggr]^{1/n}
\Biggl[ \frac{(n+1)L}{r}dr^2 + 2Lr d\Sigma_n^2
+ \frac{Lr}{n+1}\left( d\chi + \omega_n\right)^2\Biggr].
\end{eqnarray}
Introducing the new radial coordinate $R$ defined by 
\begin{eqnarray}
r=\frac{1}{4(n+1)L}\left( R^{2n} - R_{\rm H}^{2n} \right)^{1/n},
\label{eq:rho}
\end{eqnarray}
we find
\begin{eqnarray}
ds^2 \simeq - \left( 1 - \frac{R_{\rm H}^{2n}}{R^{2n}}\right)^2 dt^2
+ \left( 1 - \frac{R_{\rm H}^{2n}}{R^{2n}}\right)^{-2} dR^2 
+ R^2 d\Omega_{2n+1}^2,
\label{eq:Dmetricrsmall}
\end{eqnarray}
where the constant $R_{\rm H}$ is given by
\begin{eqnarray}
R_{\rm H} = \sqrt{2(n+1)} \left( \frac{n+1}{n} \right)^{1/(2n)} \mu.
\label{eq:horizonradius}
\end{eqnarray}
The metrics (\ref{eq:Dmetricrsmall}) are the $D$-dimensional extremal 
Reissner-Nordstr\"{o}m metrics \cite{Tangherlini:1963bw}.

In the limit $r\to 0$, i.e., $R \to R_{\rm H}$, 
the metrics (\ref{eq:Dmetricrsmall})  behave as 
\begin{eqnarray}
ds^2 \to - \Biggl[ \frac{2n(R-R_{\rm H})}{R_{\rm H}} \Biggr]^2 dt^2 
+ \Biggl[ \frac{2n(R-R_{\rm H})}{R_{\rm H}} \Biggr]^{-2}dR^2 
+ R_{\rm H}^2 d\Omega_{2n+1}^2 .
\end{eqnarray}
The metrics are rewritten as
\begin{eqnarray}
ds^2 \simeq - \exp \left(\frac{4n{\tilde r}}{R_{\rm H}}\right) d{\tilde t}^2
+d{\tilde r}^2 + R_{\rm H}^2 d\Omega_{2n+1}^2,
\label{eq:Dmetricr0}
\end{eqnarray}
by using the new coordinates ${\tilde r}$ and ${\tilde t}$ given by
\begin{eqnarray}
R = R_{\rm H}\Biggl[ \exp \left( \frac{2n {\tilde r}}{R_{\rm H}} \right) 
+ 1 \Biggr] 
, \quad 2nt = {\tilde t}.
\end{eqnarray}
The metrics \eqref{eq:Dmetricr0} imply the well known fact that 
the extremal Reissner-Nordstr\"{o}m metrics approaches
the metrics of ${\rm AdS_2 \times S^{2n+1}}$ at the horizon \cite{Reall:2002bh,Kunduri:2007vf}.
Therefore, one might expect that the solution (\ref{eq:Dmetric}) is regular at $r=0$.
Indeed, in the $D=5$ case, the extremal charged Kaluza-Klein black hole with a 
twisted ${\rm S^1}$ is smooth at the horizon~\cite{Kimura:2008cq}. 
However, higher-order terms that we ignore cause the singularity in $D>5$, 
as will be shown later. 

If $r \gg L, \mu$, the metrics 
behave as
\begin{eqnarray}
ds^2_{D=5}&\simeq& -\left( 1-\frac{2\mu^2}{Lr}\right) dt^2
+\left( 1+\frac{2L^2+\mu^2}{Lr}\right) dr^2 
+	r^2\left( 1+\frac{2L^2+\mu^2}{Lr}\right) d\Sigma_1^2 \nonumber 
\\
&& + L^2\left( 1-\frac{2L^2-\mu^2}{Lr}\right) \left( d\chi +\omega_1\right)^2
\label{eq:5metricrlarge}
\end{eqnarray}
for $D=5$, and
\begin{eqnarray}
ds^2_{D>5} &\simeq& -\left( 1-\frac{2\mu^{2n}}{Lr^{2n-1}}\right) dt^2 
+ (2n-1) \left( 1 + \frac{2n}{2n-3}\frac{L^2}{r^2}\right)
dr^2 + r^2 \left( 1+\frac{2L}{r}\right) d\Sigma_n^2 \nonumber 
\\
&&	+ \frac{1}{2n-1}L^2\left( 1-\frac{2n}{2n-3}\frac{L^2}{r^2}\right) 
\left( d\chi +\omega_n\right)^2
\label{eq:Dmetricrlarge}
\end{eqnarray}
for $D>5$. 
In Eqs.(\ref{eq:5metricrlarge}) and (\ref{eq:Dmetricrlarge}), 
since the coordinate $r$ is the circumferential radius with respect to  
$\mathbb{C}P(n)$, in this coordinate, $g_{tt}$ gives 
the \lq Newtonian gravitational potential\rq\ in the large scale.
The $r$ dependences of $g_{tt}$ imply the gravity in the far region is 
dimensionally reduced. 
Note that one can write $g_{tt}$ with the $D$-dimensional gravitational constant $G_D$ explicitly in the form
\begin{eqnarray}
-g_{tt} = 1 - \frac{2G_{D} \mu^{2n} }{L r^{2n-1} },  
\end{eqnarray}
where $G_D/L$ is the $(D-1)$-dimensional gravitational constant. 

\section{Properties of $r=0$ Surface}
\label{sec:4}

In the $\mu> 0$ case, the $r=0$ surface is an apparently singular null surface 
in the metric form 
\eqref{eq:Dmetric} with \eqref{eq:H}. 
So,  we first 
construct coordinates which cover the surface $r = 0$
to see the smoothness of this surface.

To avoid the coordinate singularity in the metric (\ref{eq:Dmetric}) at $r=0$, 
we introduce the coordinates $(v, \, \rho )$: 
\begin{eqnarray}
dv &=& dt + H^{1+1/n}F^{-1/2}dr, \\
d\rho &=& \frac{n L^n r^{n-1}}{R_{\rm H}^{2n-1}}dr.
\label{eq:dzeta}
\end{eqnarray}
Eq.(\ref{eq:dzeta}) means $\rho \propto r^{n}$. 
Then, the metric (\ref{eq:Dmetric}) becomes
\begin{eqnarray}
ds^2 &=& - H^{-2}dv^2 
+ H^{-1+1/n}F^{-1/2} \frac{R_{\rm H}^{2n-1}}{n L^n r^{n-1}} dv d\rho \nonumber \\
&& + H^{1/n}[ r(r+2L)d\Sigma_n^2 + L^2 F(d\chi + \omega_n )^2 ],
\label{eq:Dmetriczeta}
\end{eqnarray}
where $r$ is a function of $\rho$.
In the case $D=5$, the spacetimes with the metrics \eqref{eq:Dmetriczeta} give an analytic extension of \eqref{eq:Dmetric} and 
the metric components in Eq.(\ref{eq:Dmetriczeta}) are smooth functions of $r$ or $\rho$ at $r=0$ as follows
\begin{eqnarray}
g^{D=5}_{v\rho} &=& 4 + \left( \frac{1}{L} - \frac{16L}{R_{\rm H}^2} \right) r + O(r^2), \nonumber \\
&=& 4 + \left( \frac{1}{L} - \frac{16L}{R_{\rm H}^2} \right) \frac{R_{\rm H}}{L}\rho + O(\rho^2), \\ 
g^{D=5}_{\xi_1 \xi_1} &=& R_{\rm H}^2 + \left( 8L + \frac{R_{\rm H}^2}{2L} \right) r + O(r^2), \nonumber \\
&=& R_{\rm H}^2 + \left( 8L + \frac{R_{\rm H}^2}{2L} \right) \frac{R_{\rm H}}{L} \rho + O(\rho^2 ).
\end{eqnarray}
However, in the case $D>5$, the spacetimes with the metrics \eqref{eq:Dmetriczeta} do not give an analytic extension 
because the metric components are not smooth functions of $\rho$ as follows
\begin{eqnarray}
g^{D>5}_{v\rho} &=& \frac{2^{n+1} (n+1)^n }{n} + \frac{2^n (n+1)^{n+1} }{n+2} \frac{r}{L} + O(r^2), \nonumber \\
&=& \frac{2^{n+1} (n+1)^n }{n} + \frac{2^n (n+1)^{n+1} }{ n+2 } \frac{R_{\rm H}^{(2n-1)/n} \rho^{1/n} }{L^2} + O(\rho^{2/n} ), \\
g^{D>5}_{\xi_n \xi_n} &=& R_{\rm H}^2 + \frac{R_{\rm H}^2 r}{(n+2)L} + O(r^2), \nonumber \\
&=& R_{\rm H}^2 + \frac{R_{\rm H}^{2 + (2n-1)/n} \rho^{1/n}}{(n+2)L^2} + O(\rho^{2/n}). 
\end{eqnarray}
We can see that the essential reason for analyticity breaking is the fractional powers of $\rho$. 
Though we can remove the fractional powers in $g_{v\rho}$ by transforming $\rho$ to a new radial coordinate but 
cannot do that in $g_{\xi_n \xi_n}$. 
This fact implies that the null hypersurface $r=0$ is a singular hypersurface. 

To examine the properties of the hypersurface $r=0$
more carefully, we use vielbeins $e^{(\alpha)}_{\mu}dx^\mu$ satisfying
\begin{eqnarray}
e^{(0)\mu}\nabla_{\mu}e^{(\alpha)\nu}=0, \ \ g^{\mu\nu}e^{(\alpha)}_{\mu}e^{(\beta)}_{\nu}
=\eta^{(\alpha)(\beta)},
\end{eqnarray}
where indices with parentheses 
distinguish each vielbein, and 
\begin{eqnarray}
\eta_{(\alpha)(\beta)} = \eta^{(\alpha)(\beta)} = {\rm diag} (-1, 1, \cdots, 1).
\end{eqnarray}
In particular, we take 
\begin{eqnarray}
e^{(0)}_\mu dx^\mu &=& -dt - H^{1/(2n)} \sqrt{\frac{H^2 - 1}{F}} dr,
\label{eq:vielbein1} \\
e^{(1)}_\mu dx^\mu &=& - \frac{\sqrt{H^2 - 1}}{H}dt - \frac{H^{1+1/(2n)}}{\sqrt{F}}dr, \\
e^{(i)}_\mu dx^\mu &=& \sqrt{2(n+1)r(r+2L)H^{1/n}} \prod_{I=i}^{n} \sin \xi_I \, d\xi_i, \ \ ({\rm for} \ i = 2, \cdots n) \\
e^{(n+1)}_\mu dx^\mu &=& \sqrt{2(n+1)r(r+2L)H^{1/n}} \, d\xi_n,
\label{eq:vielbein3} \\
e^{(j)}_\mu dx^\mu &=& \frac{1}{2(j-n)}\sqrt{2(n+1)r(r+2L)H^{1/n}} \cot \xi_{j-n-1} \prod_{I=j-n}^{n} \sin \xi_I \, \omega_{j-n-1}, \\
&& ({\rm for} \ j = n+2, \cdots 2n) \nonumber \\
e^{(2n+1)}_\mu dx^\mu &=& \frac{1}{2(n+1)}\sqrt{2(n+1)r(r+2L)H^{1/n}} \cot \xi_{n} \, \omega_n, \\
e^{(2n+2)}_\mu dx^\mu &=& L\sqrt{F H^{1/n}} (d\chi + \omega_n),
\label{eq:vielbein2}
\end{eqnarray}
where $e^{(0)\mu}\partial_{\mu}$ is chosen as the geodesic tangent vector 
in the radial direction.

First, we calculate the expansion for an outgoing null geodesic congruence $k^\mu \partial_\mu$
defined by
\begin{eqnarray}
\theta_+ &=& h^{\mu\nu}\nabla_{\mu}k_{\nu},
\label{eq:theta} \\
h^{\mu\nu} &=& g^{\mu\nu} + k^\mu l^\nu + k^\nu l^\mu, 
\label{eq:h}
\end{eqnarray}
where $k^\mu k_\mu = l^\mu l_\mu = 0$ and $k^\mu l_\mu = -1$. 
Using the vielbein frame, 
we introduce outgoing and ingoing null vector fields $k^\mu \partial_\mu$ and 
$l^\mu \partial_\mu$ by
\begin{eqnarray}
k^\mu \frac{\partial}{\partial x^\mu} &=& \frac{1}{\sqrt{2}} \left( e^{(0)\mu} - e^{(1)\mu} \right) \partial_{\mu} \nonumber \\
&=& \frac{H(H-\sqrt{-1+H^2})}{\sqrt{2}}\frac{\partial}{\partial t}
+ \frac{\sqrt{F}(H-\sqrt{-1+H^2})}{\sqrt{2}H^{1/(2n)}}\frac{\partial}{\partial r}, \\
l^\mu \frac{\partial}{\partial x^\mu} &=& \frac{1}{\sqrt{2}} \left( e^{(0)\mu} + e^{(1)\mu} \right) \partial_{\mu} \nonumber \\
&=& \frac{H(H+\sqrt{-1+H^2})}{\sqrt{2}}\frac{\partial}{\partial t}
- \frac{\sqrt{F}(H+\sqrt{-1+H^2})}{\sqrt{2}H^{1/(2n)}}\frac{\partial}{\partial r}. 
\end{eqnarray}
The vectors $k^\mu \partial_\mu$ are tangent to the null geodesics which are 
orthogonal to the $r={\rm const.}$ surfaces. 
Then, we find that
\begin{eqnarray}
\theta_+ &=& \frac{1-H^2+H\sqrt{-1+H^2}}{2\sqrt{2} n r(r+2L)\sqrt{F} H^{(2n+1)/(2n)}}  \nonumber \\
&& \times \Biggl[ nr(r+2L)\frac{dF}{dr}H + F\left( 4n^2 (r+L) H + (2n+1)r (r+2L) \frac{dH}{dr} \right) \Biggr].
\end{eqnarray}
In the vicinity of $r = 0$, the expansion $\theta_+$ behaves as, 
for example,  
in the $D=5, 7, 9$ cases,
\begin{eqnarray}
\theta_+^{D=5} &\simeq& \frac{r^2 (6L^2 + \mu^2 )}{16 \mu^5 }, 
\label{eq:theta5} \\
\theta_+^{D=7} &\simeq& \frac{r^4 [320L^4 - 36\mu^4 + 15\mu^4 \ln (4L/r) ]}{18 \times 6^{3/4} \mu^9 }, 
\label{eq:theta7} \\
\theta_+^{D=9} &\simeq& \frac{63 \times 3^{1/6} L r^5}{100 \times 2^{1/3} \mu^7 } \nonumber \\ 
&& + \frac{9 \times 3^{1/6} r^6 [ 7000 L^6 + 800 \mu^6 + 49 \pi \mu^6 -98 \mu^6 \tan^{-1}3 - 336 \mu^6 \ln (\sqrt{10}L/r) ] }
{2000 \times 2^{1/3} \times \mu^{13} }. 
\label{eq:theta9}
\end{eqnarray}
Since the expansion $\theta_+^{D=2n+3}$ is positive in the region $r>0$ and zero at the null hypersurface $r=0$, 
the hypersurface $r=0$ looks like the apparent horizon for observers in the region $r>0$. 
In the limit $L \to \infty$, 
the expansion behaves as
\begin{eqnarray}
\theta_+^{D=2n+3} \simeq \frac{(2n+1)[4(n+1)Lr]^{2n}}{2\sqrt{2} R_{\rm H}^{4n+1}} = \frac{(2n+1)(R^{2n} - R_{\rm H}^{2n})^2}{2\sqrt{2} R_{\rm H}^{4n+1}}. 
\label{eq:thetaD}
\end{eqnarray}
This reproduces the expansion for $r={\rm const.}$ in the case of the Reissner-Nordstr\"om spacetimes. 
Note that the leading behavior of $\theta_+$ near $r=0$ is different from \eqref{eq:thetaD}, though the leading behavior of the metric becomes Reissner-Nordstr\"om spacetime
as shown in Sec.III.B.  

While we can find an analytic extension across the horizon $r=0$ for $D=5$, 
but hardly find such an extension for $D>5$.
Although we can show that the Kretschmann invariant 
$R_{\mu\nu\rho\sigma}R^{\mu\nu\rho\sigma}$ is finite at $r=0$ for all $D$, 
but cannot conclude that the horizon $r=0$ is smooth.  
To examine the regularity at $r=0$, 
we consider
the Riemann curvature measured by a free-falling observer with orthonormal bases 
\eqref{eq:vielbein1}-\eqref{eq:vielbein2}~\cite{Horowitz:1997si}. 

We 
can calculate the vielbein components of the Riemann curvature, and show 
the components explicitly, for example, 
\begin{eqnarray}
R^{(0)(2)(0)(2)} = -\frac{F H^{-2-1/n}(-1+H^2)}{4n^2 r^2 (r+2L)^2} \Biggl[2n(r+L)H+r(r+2L) \left(\frac {dH}{dr} \right) \Biggr]^2.
\end{eqnarray}
In the limit $r \to 0$, the 
above curvature component behaves as
\begin{eqnarray}
R^{(0)(2)(0)(2)}_{D=5} &\simeq& -\frac{(2L^2 + \mu^2 )^2}{32L^4 \mu^2} 
\label{eq:riemann5}
\end{eqnarray}
for $D=5$, and
\begin{eqnarray}
R^{(0)(2)(0)(2)}_{D>5} &\propto& -\frac{\mu^{4n-2}}{L^{2(n+1)}r^{2(n-1)}}. 
\label{eq:riemannD}
\end{eqnarray}
for $D>5$. 
While this component of the Riemann curvature is finite for $D=5$, 
the Riemann curvature diverges at $r=0$ for $D>5$.  
Then, the solutions are singular at $r=0$ for $D>5$, 
and there is no C$^2$ extension across the $r=0$ surface. 
This makes a contrast with the case of $D=5$.
It is consistent that the singularity at $r=0$ disappears in the limit $L \to \infty$ with the case of the Reissner-Nordstr\"om metrics. 
As shown in Appendix A, there exists a C$^{0}$ extension across the horizon, 
which is consistent that Eq.\eqref{eq:riemannD} can be rewritten as $\rho^{-2+1/n}$ with the regular coordinate $\rho$. 
Furthermore, we can see that a tidal force causes finite difference in deviation of free-fall geodesic 
congruences just before and after crossing the horizon. 

\section{solutions with a positive cosmological constant}

In this section, we consider the action with a cosmological constant $\Lambda$,
\begin{eqnarray}
S = \frac{1}{16\pi} \int d^D x \sqrt{-g} \left( R - F_{\mu\nu}F^{\mu\nu} 
- 4\Lambda \right).
\end{eqnarray}
The Einstein and Maxwell equations are
\begin{eqnarray}
R_{\mu\nu} - \frac{1}{2}R g_{\mu\nu} + 2\Lambda g_{\mu\nu} &=& 2 \left( F_{\mu\alpha}F_{\nu}^{\ \alpha} - \frac{1}{4}F_{\alpha\beta}F^{\alpha\beta}g_{\mu\nu} \right), 
\label{eq:Einsteinlambda} \\
\nabla^{\mu}F_{\mu\nu} &=& 0.
\label{eq:Maxwelllambda}
\end{eqnarray}

We can construct the $D=(2n+3)$-dimensional cosmological black holes 
on generalized Taub-NUT spaces, 
if the cosmological constant $\Lambda$ is positive, 
using harmonic functions in the similar way in Sec.II. 
The metrics and the Maxwell fields
\begin{eqnarray}
ds^2 &=& - H(t,r)^{-2}{dt^2} \nonumber \\
&& + [ H(t,r) e^{-\lambda t}]^{1/n} 
\Biggl[ \frac{dr^2}{F(r)} + r(r+2L)d\Sigma_n^2
+ L^2 F(r) (d\chi + \omega_n)^2 \Biggr],  
\label{eq:cosDmetric} \\
A_\mu dx^{\mu} &=& \pm \sqrt{\frac{2n+1}{4n}} \frac{dt}{H(t,r)}, 
\end{eqnarray}
solve Eqs. \eqref{eq:Einsteinlambda} and \eqref{eq:Maxwelllambda}, 
where
\begin{eqnarray}
H(t,r)&=&1+\frac{\mu^{2n}}{e^{-\lambda t} } 
\int_{r}^{\infty} dr \frac{1}{L[r(r+2L)]^{n}F(r)}, \\
\lambda &=& \sqrt{\frac{8n^2\Lambda}{(n+1)(2n+1)}} > 0,	
\end{eqnarray}
and $F(r)$ is given by Eq.(\ref{eq:F}).

To examine where the horizon exists, 
we consider the expansion $\theta_+$ 
for a congruence of outgoing null geodesics emanating from an $r={\rm const.}$ sphere 
on a time slice. 
We take the outgoing and ingoing null vector fields perpendicular 
to the sphere be
\begin{eqnarray}
k^{\mu}\frac{\partial}{\partial x^{\mu}} 
&=& \frac{1}{\sqrt{2}} H(t,r)~ \frac{\partial}{\partial t} 
+ \sqrt{\frac{F(r)}{2 [e^{-\lambda t}H(t,r)]^{1/n}}}~ 
\frac{\partial}{\partial r}, \\
l^{\mu}\frac{\partial}{\partial x^{\mu}} 
&=& \frac{1}{\sqrt{2}} H(t,r)~ \frac{\partial}{\partial t} 
- \sqrt{\frac{F(r)}{2 [e^{-\lambda t}H(t,r)]^{1/n}}}~ 
\frac{\partial}{\partial r}, 
\end{eqnarray}
so that $k^\mu l_\mu =-1$. 
We find that the expansion for the outgoing null geodesic congruence, defined by Eqs.(\ref{eq:theta}) and (\ref{eq:h}), is in the form
\begin{eqnarray}
\theta_+ &=& \frac{1}{2\sqrt{2} F^{1/2}(e^{-\lambda t}H)^{1/(2n)}} \frac{\partial F}{\partial r}
+\frac{2n(r+L)F^{1/2}}{\sqrt{2} r(r+2L)(e^{-\lambda t}H)^{1/(2n)}} 
- \frac{(2n+1)\lambda H}{2\sqrt{2} n} \nonumber \\
&& + \frac{(2n+1)F^{1/2}}{2\sqrt{2} n(e^{-\lambda t}H)^{1/(2n)}}\frac{1}{H} \frac{\partial H}{\partial r} 
+ \frac{(2n+1)}{2\sqrt{2} n} \frac{\partial H}{\partial t}. 
\label{exact_expansion}
\end{eqnarray}
Using the new variable $R$ defined by 
\begin{eqnarray}
R^{2n} - R_{\rm H}^{2n} 
= [4(n+1)Lr]^{n} e^{-\lambda t}, 
\label{eq:R_trho_tr}
\end{eqnarray}
we rewrite \eqref{exact_expansion} approximately in the region $r \ll L$ as
\begin{eqnarray}
\theta_+ \simeq \frac{2n+1}{\sqrt{2}} \left( \frac{1}{R} 
- \frac{R_{\rm H}^{2n}}{R^{2n+1}} - \frac{\lambda}{2n} \right)
\label{eq:thetaDlambda}
\end{eqnarray}
as a function of $R$.
This coincides with the expansion for $r={\rm const.}$ in the case of Reissner-Nordstr\"om-de Sitter spacetimes. 
The equation $\theta_+(R) = 0$ given in \eqref{eq:thetaDlambda} has two roots $R_{\rm BH}$ and $R_{\rm CH}$ ($R_{\rm BH} < R_{\rm CH} $)
in the region $R>R_{\rm H}$
if the condition 
\begin{eqnarray}
0 < (\lambda R_{\rm H})^{2n} < \frac{(2n)^{4n}}{(2n+1)^{2n+1}}
\end{eqnarray}
is satisfied (see Appendix B).
The large one corresponds to the cosmological horizon, and the small one to the 
apparent horizon of the black hole. 
According to \eqref{eq:R_trho_tr}, the location of the apparent horizon written by $t$ and $r$ is in the form
\begin{eqnarray}
r = e^{\lambda t/n}\frac{(R_{\rm BH}^{2n}-R_{\rm H}^{2n})^{1/n}}{4(n+1)L}. 
\end{eqnarray}
Because 
we have assumed $r \ll L$, it is necessary that 
$e^{\lambda t}$ 
is small, i.e., we should consider 
an early stage $t \to -\infty$.
In this stage, we can find the following facts. 
First, the metric components at the apparent horizon are smooth, and the $r=0$ surface is hidden in the trapped region. 
Secondly, the geometry of these spacetimes near the horizon approaches to that of Reissner-Nordstr\"om-de Sitter spacetimes with $M=|Q|\sqrt{(2n+1)/(4n)}$
because the size of the compactified extra dimension is very large. 

\section{Summary and discussions}

Focusing on odd dimensions, 
we have obtained extremal charged static exact solutions 
in the $D$-dimensional Einstein-Maxwell theory 
using higher-dimensional generalizations of the four-dimensional Euclidean Taub-NUT space. 
These solutions asymptote to $(D-1)$-dimensional spacetimes 
with a twisted $\rm S^1$ as an extra dimension, 
where the angular parts of the $(D-1)$-dimensional spacetime are $\mathbb{C}P(n)$. 
In the $D = 5$ case, the solution
coincides with the extremal charged Kaluza-Klein black hole solution 
\cite{
Ishihara:2006iv, Gauntlett:2002nw,Gaiotto:2005gf,Matsuno:2008fn,Bena:2009ev}.

In the both cases $D=5$ and $D>5$, 
there are null hypersurfaces, 
where the expansion of an outgoing null geodesic congruence 
vanishes. 
These spacetimes look like black holes for an observer 
outside the hypersurface. 
We have calculated the Riemann curvature in a frame of 
an observer parallelly transported along a free-fall geodesic. 
As a result, there exists a qualitative difference 
between the $D = 5$ and the $D > 5$ cases. 
In the $D = 5$ case, 
the solution describes a regular black hole, 
where an analytic extension across the horizon exists \cite{Kimura:2008cq}. 
In contrast, in the $D>5$ case, though Kretschmann invariant is finite\footnote{We verified the finiteness of Kretschmann invariants up to in eleven dimensions with Mathematica.}, the solution has the curvature singularity 
at the null hypersurface. 

We can find C$^0$ extension across the horizon $r=0$ and the observer along the free-fall geodesic can traverse the horizon.  
This is because the singularity is relatively mild, i.e., a tidal force causes finite difference in deviation of geodesic congruence just before and after crossing the horizon. 

We have generalized the solutions 
to the case with a positive cosmological constant.
We have obtained extremal charged solutions which have smooth 
apparent horizons in the very early stage at least
as same as the extremal Reissner-Nordstr\"om black holes 
in the asymptotically de Sitter space \cite{Brill:1993tw}.    

\section*{Acknowledgments}
The authors would like to thank T. Houri, 
K. Nakao, Y. Yasui, and the members of the YITP cosmology group 
for useful discussions.
This work is supported by the Grant-in-Aid for Scientific Research No.19540305. 
MK is supported by the JSPS Grant-in-Aid for Scientific Research No.11J02182.

\appendix
\section{C$^0$ extension across the horizon $r=0$}
We can construct the coordinates across the horizon as follows.  
We introduce a new coordinate $\zeta$ given by
\begin{eqnarray}
\frac{1}{L^{2n-1} \zeta} = \frac{H(r) - 1}{\mu^{2n}}.
\end{eqnarray}
Then, the metric (\ref{eq:Dmetric}) is represented by
\begin{eqnarray}
ds^2 &=& - \left( 1 + \frac{m}{\zeta} \right)^{-2} dt^2 \nonumber \\
&& + \left( 1 + \frac{m}{\zeta} \right)^{1/n} 
\Biggl[ \frac{1}{F} \left( \frac{dH}{dr} \right)^{-2} \frac{m^2}{\zeta^4} d\zeta^2 + r(r+2L)d\Sigma_n^2 + L^2 F (d\chi + \omega_n )^2 \Biggr], 
\label{eq:Cmetric}
\end{eqnarray}
where $m = \mu^{2n}/L^{2n-1}$. 
To extend the solutions across $\zeta=0$, we introduce null coordinate $v$ such as
\begin{eqnarray}
dv = dt + \left( 1 + \frac{m}{\zeta} \right)^{1 + 1/(2n)} \frac{1}{F^{1/2}} \left( \frac{dH}{dr} \right)^{-1} \frac{m}{\zeta^2} d\zeta.
\end{eqnarray}
Then, the metric (\ref{eq:Cmetric}) becomes
\begin{eqnarray}
ds^2 &=& - \left( 1 + \frac{m}{\zeta} \right)^{-2} dv^2 + 2 \left( \zeta + m \right)^{-1 + 1/(2n)} \frac{1}{\zeta^{-1+1/(2n)}} \frac{1}{F^{1/2}} 
\left( \frac{dH}{dr} \right)^{-1} \frac{m}{\zeta^2} dv d\zeta
\nonumber \\
&& + \left( \zeta + m \right)^{1/n} \frac{1}{\zeta^{1/n}} \Bigl[ r(r+2L)d\Sigma_n^2 + L^2 F (d\chi + \omega_n)^2 \Bigr]. 
\label{eq:Cmetric2}
\end{eqnarray}
We assume that this metric extended in the region $\zeta<0$ as follows:
\begin{eqnarray}
ds^2 &=& - \left( 1 + \frac{m}{\zeta} \right)^{-2} dv^2 + 2 \left( \zeta + m \right)^{-1 + 1/(2n)} 
\frac{1}{|\zeta|^{-1+1/(2n)}} \frac{1}{F^{1/2}} \left( \frac{dH}{dr} \right)^{-1} \frac{m}{\zeta^2} dv d\zeta
\nonumber \\
&& + \left( \zeta + m \right)^{1/n} \frac{1}{|\zeta|^{1/n}} \Bigl[ r(r+2L)d\Sigma_n^2 + L^2 F (d\chi + \omega_n)^2 \Bigr], 
\label{eq:Cmetric4}
\end{eqnarray}	
where $r$ is the function of $\zeta$ implicitly represented by
\begin{eqnarray}
\frac{1}{L^{2n-1} |\zeta|} = \frac{H(r) - 1}{\mu^{2n}}.
\end{eqnarray}
Introducing a coordinate ${\bar \zeta} = - \zeta$, we show that in the region $\zeta < 0$, the above metric becomes
\begin{eqnarray}
ds^2 &=& - \left( -1 + \frac{m}{{\bar \zeta}} \right)^{-2} dv^2 
- 2 \left( -1 + \frac{m}{{\bar \zeta}} \right)^{-1 + 1/(2n)} \frac{1}{F^{1/2}} \left( \frac{dH}{dr} \right)^{-1} \frac{m}{{\bar \zeta}^2} dv d{\bar \zeta}
\nonumber \\
&& + \left( -1 + \frac{m}{{\bar \zeta}} \right)^{1/n} \Bigl[ r(r+2L)d\Sigma_n^2 + L^2 F (d\chi + \omega_n)^2 \Bigr]. 
\label{eq:Cmetric3}
\end{eqnarray}
The metric in above form is obtained by replacing the constant of integration $1$ with $-1$ and `time' reversal in (\ref{eq:Cmetric2}). 
Therefore, it is clear that this metric is the solution of the Einstein-Maxwell equations. 
The spacetimes with the metrics \eqref{eq:Cmetric4} give a C$^0$ extension for the spacetimes with the metrics \eqref{eq:Cmetric}.
		
\section{Reissner-Nordstr\"{o}m-de Sitter spacetime with $M=|Q|\sqrt{(2n+1)/(4n)}$}
We briefly review the $(2n+3)$-dimensional Reissner-Nordstr\"{o}m-de Sitter (RNdS) spacetime with
$M=|Q|\sqrt{(2n+1)/(4n)}$, 
where $M$ and $Q$ are Komar mass and charge respectively.
The metrics and the Maxwell fields are written in static coordinates as
\begin{eqnarray}
ds^2 &=& -f(R)dT^2 + f(R)^{-1}dR^2 + R^2 d\Omega_{2n+1}^2, \\
A_{\mu} dx^{\mu} &=& \pm \sqrt{\frac{2n+1}{4n}} \left( 1 - \frac{R_{\rm H}^{2n}}{R^{2n}} \right) dT,
\end{eqnarray}
where 
\begin{eqnarray}
f(R) = \left( 1 - \frac{R_{\rm H}^{2n}}{R^{2n}} \right)^2 - \frac{\lambda ^2 R^2}{4n^2}. 
\end{eqnarray}
The parameter $R_{\rm H}$ is related to the mass $M$ as
\begin{eqnarray}
R_{\rm H}^{2n} = \frac{8\pi}{(2n+1) \Omega_{2n+1}} M. 
\end{eqnarray}
Introducing the cosmological coordinates $(r, t)$ in the form~\cite{London:1995ib}
\begin{eqnarray}
e^{-\lambda t}r^{2n} &=& R^{2n} - R_{\rm H}^{2n}, \ \ \ \ t = T + W (R), \\
\frac{dW (R)}{dr} &=& \frac{\lambda R^{2n+1}}{2n (R^{2n} - R_{\rm H}^{2n})} \frac{1}{f(R)},
\end{eqnarray}
we find
\begin{eqnarray}
ds^2 = - \left( 1 + \frac{R_{\rm H}^{2n}}{e^{-\lambda t}r^{2n}} \right)^{-2}dt^2 
+ \left( 1 + \frac{R_{\rm H}^{2n}}{e^{-\lambda t}r^{2n}} \right)^{1/n} e^{-\lambda t/n} (dr^2 + r^2 d\Omega_{2n+1}^2).
\end{eqnarray}
We assume that outgoing and ingoing null vector fields are
\begin{eqnarray}
k^{\mu}\frac{\partial}{\partial x^{\mu}} &=& \frac{1}{\sqrt{2}} 
\left( 1 + \frac{R_{\rm H}^{2n}}{e^{- \lambda t}r^{2n}} \right)
\frac{\partial}{\partial t} 
+ \frac{1}{\sqrt{2}} 
e^{\lambda t/2n}\left( 1 + \frac{R_{\rm H}^{2n}}{e^{- \lambda t}r^{2n}} \right)^{-1/(2n)} 
\frac{\partial}{\partial r}, \\
l^{\mu}\frac{\partial}{\partial x^{\mu}} &=& \frac{1}{\sqrt{2}} 
\left( 1 + \frac{R_{\rm H}^{2n}}{e^{- \lambda t}r^{2n}} \right) 
\frac{\partial}{\partial t} 
- \frac{1}{\sqrt{2}} 
e^{\lambda t/2n}\left( 1 + \frac{R_{\rm H}^{2n}}{e^{- \lambda t}r^{2n}} \right)^{-1/(2n)}
\frac{\partial}{\partial r}.
\end{eqnarray}
Then, we find
\begin{eqnarray}
\theta_+ = && - \frac{2n+1}{2\sqrt{2}n} ( e^{- \lambda t}r^{2n} )^{-2}
\left( 1 + \frac{R_{\rm H}^{2n}}{e^{- \lambda t}r^{2n}} \right)^{-2}
\Biggl[ \lambda R_{\rm H}^{4n} + 2\lambda R_{\rm H}^{2n} ( e^{- \lambda t}r^{2n} ) \nonumber \\
&& - 2n \left( 1 + \frac{R_{\rm H}^{2n}}{e^{- \lambda t}r^{2n}}\right)^{(2n-1)/(2n)} ( e^{- \lambda t}r^{2n} )^{(4n-1)/(2n)}
+ \lambda ( e^{- \lambda t}r^{2n} )^{2} \Biggr].
\label{eq:expansionr}
\end{eqnarray}
Using the static coordinates $(T, R)$, we can write $\theta_+$ in the form
\begin{eqnarray}
\theta_+ = \frac{2n+1}{\sqrt{2}} \left( \frac{1}{R} - \frac{R_{\rm H}^{2n}}{R^{2n+1}} - \frac{\lambda}{2n} \right).
\label{eq:expansionR}
\end{eqnarray}
The condition that the equation $\theta_+ = 0$ has two positive roots $R_{\rm BH}$ and $R_{\rm CH}$ ($R_{\rm H} < R_{\rm BH} < R_{\rm CH}$) is~\cite{Kodama:2003kk}
\begin{eqnarray}
0 < (\lambda R_{\rm H})^{2n} < \frac{(2n)^{4n}}{(2n+1)^{2n+1}}.
\label{eq:lambdacondition}
\end{eqnarray}
The small one corresponds to the black hole horizon and the large one to the cosmological horizon. 



\begin{thebibliography}{999}

\bibitem{Emparan:2001wn}
  R.~Emparan and H.~S.~Reall,
  Phys.\ Rev.\ Lett.\  {\bf 88}, 101101 (2002)
  [arXiv:hep-th/0110260].

\bibitem{Tangherlini:1963bw}
  F.~R.~Tangherlini,
  Nuovo Cim.\  {\bf 27}, 636-651 (1963).
  
\bibitem{Myers:1986un}
  R.~C.~Myers and M.~J.~Perry,
  Annals Phys.\  {\bf 172}, 304 (1986).


\bibitem{Mishima:2005id}
  T.~Mishima and H.~Iguchi,
  Phys.\ Rev.\  D {\bf 73}, 044030 (2006)
  [arXiv:hep-th/0504018].

\bibitem{Figueras:2005zp}
  P.~Figueras,
  JHEP {\bf 0507}, 039 (2005)
  [arXiv:hep-th/0505244].

\bibitem{Pomeransky:2006bd}
  A.~A.~Pomeransky and R.~A.~Sen'kov,
  arXiv:hep-th/0612005.

\bibitem{Elvang:2007rd}
  H.~Elvang and P.~Figueras,
  JHEP {\bf 0705}, 050 (2007)
  [arXiv:hep-th/0701035].

\bibitem{Iguchi:2007is}
  H.~Iguchi and T.~Mishima,
  Phys.\ Rev.\  D {\bf 75}, 064018 (2007)
  [arXiv:hep-th/0701043].

\bibitem{Izumi:2007qx}
  K.~Izumi,
  Prog.\ Theor.\ Phys.\  {\bf 119}, 757 (2008)
  [arXiv:0712.0902 [hep-th]].

\bibitem{Elvang:2007hs}
  H.~Elvang and M.~J.~Rodriguez,
  JHEP {\bf 0804}, 045 (2008)
  [arXiv:0712.2425 [hep-th]].

\bibitem{Evslin:2008py}
  J.~Evslin and C.~Krishnan,
  JHEP {\bf 0809}, 003 (2008)
  [arXiv:0804.4575 [hep-th]].

\bibitem{Dobiasch:1981vh}
  P.~Dobiasch and D.~Maison,
  Gen.\ Rel.\ Grav.\  {\bf 14}, 231 (1982).

\bibitem{Gibbons:1985ac}
  G.~W.~Gibbons and D.~L.~Wiltshire,
  Annals Phys.\  {\bf 167}, 201 (1986)
  [Erratum-ibid.\  {\bf 176}, 393 (1987)].
  
\bibitem{Gauntlett:2002nw}
  J.~P.~Gauntlett, J.~B.~Gutowski, C.~M.~Hull, S.~Pakis and H.~S.~Reall,
  Class.\ Quant.\ Grav.\  {\bf 20}, 4587 (2003)
  [arXiv:hep-th/0209114].
  
\bibitem{Gaiotto:2005gf}
  D.~Gaiotto, A.~Strominger, X.~Yin,
  JHEP {\bf 0602}, 024 (2006).
  [hep-th/0503217].    

\bibitem{Ishihara:2005dp}
  H.~Ishihara and K.~Matsuno,
  Prog.\ Theor.\ Phys.\  {\bf 116}, 417 (2006)
  [arXiv:hep-th/0510094].

\bibitem{Yazadjiev:2006iv}
  S.~S.~Yazadjiev,
  Phys.\ Rev.\  {\bf D74}, 024022 (2006).
  [hep-th/0605271].  

\bibitem{Nakagawa:2008rm}
  T.~Nakagawa, H.~Ishihara, K.~Matsuno and S.~Tomizawa,
  Phys.\ Rev.\  D {\bf 77}, 044040 (2008)
  [arXiv:0801.0164 [hep-th]].

\bibitem{Tomizawa:2008hw}
  S.~Tomizawa, H.~Ishihara, K.~Matsuno and T.~Nakagawa,
  Prog.\ Theor.\ Phys.\  {\bf 121}, 823 (2009)
  [arXiv:0803.3873 [hep-th]].
  
\bibitem{Tomizawa:2008rh}
  S.~Tomizawa and A.~Ishibashi,
  Class.\ Quant.\ Grav.\  {\bf 25}, 245007 (2008)
  [arXiv:0807.1564 [hep-th]].

\bibitem{Stelea:2008tt}
  C.~Stelea, K.~Schleich and D.~Witt,
  Phys.\ Rev.\  D {\bf 78}, 124006 (2008)
  [arXiv:0807.4338 [hep-th]].
  
\bibitem{Tomizawa:2008qr}
  S.~Tomizawa, Y.~Yasui, Y.~Morisawa,
  Class.\ Quant.\ Grav.\  {\bf 26}, 145006 (2009).
  [arXiv:0809.2001 [hep-th]].  
  
\bibitem{Bena:2009ev}
  I.~Bena, G.~Dall'Agata, S.~Giusto, C.~Ruef, N.~P.~Warner,
  JHEP {\bf 0906}, 015 (2009).
  [arXiv:0902.4526 [hep-th]].

\bibitem{Tomizawa:2010xq}
  S.~Tomizawa,
  arXiv:1009.3568 [hep-th]. 

\bibitem{Mizoguchi:2011zj}
  S.~'y.~Mizoguchi, S.~Tomizawa,
  Phys.\ Rev.\  {\bf D84}, 104009 (2011).
  [arXiv:1106.3165 [hep-th]].


\bibitem{Myers:1986rx}
  R.~C.~Myers,
  Phys.\ Rev.\  {\bf D35 } (1987)  455.
  

\bibitem{Breckenridge:1996is}
  J.~C.~Breckenridge, R.~C.~Myers, A.~W.~Peet and C.~Vafa,
  Phys.\ Lett.\  B {\bf 391}, 93 (1997)
  [arXiv:hep-th/9602065].

\bibitem{Herdeiro:2002ft}
  C.~A.~R.~Herdeiro,
  Nucl.\ Phys.\  {\bf B665}, 189-210 (2003).
  [hep-th/0212002].

\bibitem{Herdeiro:2003un}
  C.~A.~R.~Herdeiro,
  Class.\ Quant.\ Grav.\  {\bf 20}, 4891-4900 (2003).
  [hep-th/0307194].

\bibitem{Brecher:2003wq}
  D.~Brecher, U.~H.~Danielsson, J.~P.~Gregory and M.~E.~Olsson,
  JHEP {\bf 0311}, 033 (2003)
  [arXiv:hep-th/0309058].
  
\bibitem{Ortin:2004af}
  T.~Ortin,
  Class.\ Quant.\ Grav.\  {\bf 22}, 939-946 (2005).
  [hep-th/0410252].  
  


\bibitem{Kim:2010bf}
  S.~S.~Kim, J.~L.~Hornlund, J.~Palmkvist and A.~Virmani,
  JHEP {\bf 1008}, 072 (2010)
  [arXiv:1004.5242 [hep-th]].

\bibitem{Elvang:2004rt}
  H.~Elvang, R.~Emparan, D.~Mateos and H.~S.~Reall,
  Phys.\ Rev.\ Lett.\  {\bf 93}, 211302 (2004)
  [arXiv:hep-th/0407065].

\bibitem{Elvang:2004ds}
  H.~Elvang, R.~Emparan, D.~Mateos and H.~S.~Reall,
  Phys.\ Rev.\  D {\bf 71}, 024033 (2005)
  [arXiv:hep-th/0408120].
  
\bibitem{Elvang:2005sa}
  H.~Elvang, R.~Emparan, D.~Mateos and H.~S.~Reall,
  JHEP {\bf 0508}, 042 (2005)
  [arXiv:hep-th/0504125].

\bibitem{Bena:2005ni}
  I.~Bena, P.~Kraus, N.~P.~Warner,
  Phys.\ Rev.\  {\bf D72}, 084019 (2005).
  [hep-th/0504142].

\bibitem{Gaiotto:2005xt}
  D.~Gaiotto, A.~Strominger, X.~Yin,
  JHEP {\bf 0602}, 023 (2006).
  [hep-th/0504126].  
  
\bibitem{Tomizawa:2007he}
  S.~Tomizawa, H.~Ishihara, M.~Kimura and K.~Matsuno,
  Class.\ Quant.\ Grav.\  {\bf 24}, 5609 (2007)
  [arXiv:0705.1098 [hep-th]].
  
\bibitem{Camps:2008hb}
  J.~Camps, R.~Emparan, P.~Figueras, S.~Giusto and A.~Saxena,
  JHEP {\bf 0902}, 021 (2009)
  [arXiv:0811.2088 [hep-th]].
  
\bibitem{Gauntlett:2004wh}
  J.~P.~Gauntlett, J.~B.~Gutowski,
  Phys.\ Rev.\  {\bf D71}, 025013 (2005).
  [hep-th/0408010].

\bibitem{Gauntlett:2004qy}
  J.~P.~Gauntlett, J.~B.~Gutowski,
  Phys.\ Rev.\  {\bf D71}, 045002 (2005).
  [hep-th/0408122].

\bibitem{Ishihara:2006pb}
  H.~Ishihara, M.~Kimura, K.~Matsuno and S.~Tomizawa,
  Phys.\ Rev.\  D {\bf 74}, 047501 (2006)
  [arXiv:hep-th/0607035].


\bibitem{Tomizawa:2008tj}
  S.~Tomizawa,
  Class.\ Quant.\ Grav.\  {\bf 25}, 145014 (2008).
  [arXiv:0802.0741 [hep-th]].




\bibitem{Maeda:2006hd}
  K.~-i.~Maeda, N.~Ohta, M.~Tanabe,
  Phys.\ Rev.\  {\bf D74}, 104002 (2006).
  [hep-th/0607084].


  
\bibitem{Ishihara:2006iv}
  H.~Ishihara, M.~Kimura, K.~Matsuno and S.~Tomizawa,
  Class.\ Quant.\ Grav.\  {\bf 23}, 6919 (2006)
  [arXiv:hep-th/0605030].



\bibitem{Matsuno:2008fn}
  K.~Matsuno, H.~Ishihara, T.~Nakagawa and S.~Tomizawa,
  Phys.\ Rev.\  D {\bf 78}, 064016 (2008)
  [arXiv:0806.3316 [hep-th]].
  

\bibitem{London:1995ib}
  L.~A.~J.~London,
  Nucl.\ Phys.\  {\bf B434}, 709-735 (1995).

\bibitem{Klemm:2000vn}
  D.~Klemm, W.~A.~Sabra,
  Phys.\ Lett.\  {\bf B503}, 147-153 (2001).
  [hep-th/0010200].
    
\bibitem{Behrndt:2004pn}
  K.~Behrndt, D.~Klemm,
  Class.\ Quant.\ Grav.\  {\bf 21}, 4107-4122 (2004).
  [hep-th/0401239].  
  
\bibitem{Gutowski:2004ez}
  J.~B.~Gutowski, H.~S.~Reall,
  JHEP {\bf 0402}, 006 (2004).
  [hep-th/0401042].

\bibitem{Gutowski:2004yv}
  J.~B.~Gutowski, H.~S.~Reall,
  JHEP {\bf 0404}, 048 (2004).
  [hep-th/0401129].

\bibitem{Ishihara:2006ig}
  H.~Ishihara, M.~Kimura, S.~Tomizawa,
  Class.\ Quant.\ Grav.\  {\bf 23}, L89 (2006).
  [hep-th/0609165].  

\bibitem{Ida:2007vi}
  D.~Ida, H.~Ishihara, M.~Kimura, K.~Matsuno, Y.~Morisawa and S.~Tomizawa,
  Class.\ Quant.\ Grav.\  {\bf 24}, 3141 (2007)
  [arXiv:hep-th/0702148].

\bibitem{Matsuno:2007ts}
  K.~Matsuno, H.~Ishihara, M.~Kimura and S.~Tomizawa,
  Phys.\ Rev.\  D {\bf 76}, 104037 (2007)
  [arXiv:0707.1757 [hep-th]].

\bibitem{Kimura:2009er}
  M.~Kimura,
  Phys.\ Rev.\  {\bf D80}, 044012 (2009).
  [arXiv:0904.4311 [gr-qc]].


\bibitem{Bais:1984xb}
  F.~A.~Bais, P.~Batenburg,
  Nucl.\ Phys.\  {\bf B253}, 162 (1985).
  
\bibitem{Awad:2000gg}
  A.~Awad, A.~Chamblin,
  Class.\ Quant.\ Grav.\  {\bf 19}, 2051-2062 (2002).
  [hep-th/0012240].

\bibitem{Cvetic:2001zb}
  M.~Cvetic, G.~W.~Gibbons, H.~Lu, C.~N.~Pope,
  Nucl.\ Phys.\  {\bf B617}, 151-197 (2001).
  [hep-th/0102185].

\bibitem{Clarkson:2002uj}
  R.~Clarkson, L.~Fatibene, R.~B.~Mann,
  Nucl.\ Phys.\  {\bf B652}, 348-382 (2003).
  [hep-th/0210280].

\bibitem{Mann:2003zh}
  R.~B.~Mann, C.~Stelea,
  Class.\ Quant.\ Grav.\  {\bf 21}, 2937-2962 (2004).
  [hep-th/0312285].

\bibitem{Lu:2004ya}
  H.~Lu, D.~N.~Page, C.~N.~Pope,
  Phys.\ Lett.\  {\bf B593}, 218-226 (2004).
  [hep-th/0403079].

\bibitem{Mann:2005gk}
  R.~B.~Mann, C.~Stelea,
  Nucl.\ Phys.\  {\bf B729}, 95-116 (2005).
  [hep-th/0505114].

\bibitem{Awad:2005ff}
  A.~M.~Awad,
  Class.\ Quant.\ Grav.\  {\bf 23}, 2849-2860 (2006).
  [hep-th/0508235].

\bibitem{Dehghani:2005zm}
  M.~H.~Dehghani, R.~B.~Mann,
  Phys.\ Rev.\  {\bf D72}, 124006 (2005).
  [hep-th/0510083].

\bibitem{Dehghani:2006aa}
  M.~H.~Dehghani, S.~H.~Hendi,
  Phys.\ Rev.\  {\bf D73}, 084021 (2006).
  [hep-th/0602069].

\bibitem{Hendi:2008wq}
  S.~H.~Hendi, M.~H.~Dehghani,
  Phys.\ Lett.\  {\bf B666}, 116-120 (2008).
  [arXiv:0802.1813 [hep-th]].

\bibitem{Hatsuda:2009vj}
  M.~Hatsuda, S.~Tomizawa,
  Class.\ Quant.\ Grav.\  {\bf 26}, 225007 (2009).
  [arXiv:0906.1025 [hep-th]].
  
\bibitem{Mann:2005ra} R.~B.~Mann and C.~Stelea,
Phys.\ Lett.\ B {\bf 634}, 448 (2006).


\bibitem{Lipert:2009sr}
  M.~Lipert,
  Class.\ Quant.\ Grav.\  {\bf 27 } (2010)  145002.
  [arXiv:0910.1170 [gr-qc]].




\bibitem{Gibbons:1994vm}
  G.~W.~Gibbons, G.~T.~Horowitz, P.~K.~Townsend,
  Class.\ Quant.\ Grav.\  {\bf 12}, 297-318 (1995).
  [hep-th/9410073].

\bibitem{Welch:1995dh}
  D.~L.~Welch,
  Phys.\ Rev.\  {\bf D52}, 985-991 (1995).
  [hep-th/9502146].

\bibitem{Candlish:2007fh}
  G.~N.~Candlish, H.~S.~Reall,
  Class.\ Quant.\ Grav.\  {\bf 24}, 6025-6040 (2007).
  [arXiv:0707.4420 [gr-qc]].

\bibitem{Candlish:2009vy}
  G.~N.~Candlish,
  Class.\ Quant.\ Grav.\  {\bf 27}, 065005 (2010).
  [arXiv:0904.3885 [hep-th]].


\bibitem{Kimura:2008cq}
  M.~Kimura,
  Phys.\ Rev.\  {\bf D78 } (2008)  047504.
  [arXiv:0805.1125 [gr-qc]].



\bibitem{Gross:1983hb}
  D.~J.~Gross, M.~J.~Perry,
  Nucl.\ Phys.\  {\bf B226}, 29 (1983).

\bibitem{Sorkin:1983ns}
  R.~d.~Sorkin,
  Phys.\ Rev.\ Lett.\  {\bf 51}, 87-90 (1983).


\bibitem{Reall:2002bh}
  H.~S.~Reall,
  Phys.\ Rev.\  {\bf D68}, 024024 (2003).
  [hep-th/0211290].



\bibitem{Kunduri:2007vf}
  H.~K.~Kunduri, J.~Lucietti and H.~S.~Reall,
  Class.\ Quant.\ Grav.\  {\bf 24}, 4169 (2007)
  [arXiv:0705.4214 [hep-th]].


\bibitem{Horowitz:1997si}
  G.~T.~Horowitz, H.~-s.~Yang,
  Phys.\ Rev.\  {\bf D55 } (1997)  7618-7624.
  [hep-th/9701077].

\bibitem{Brill:1993tw}
  D.~R.~Brill and S.~A.~Hayward,  
  Class.\ Quant.\ Grav.\  {\bf 11}, 359 (1994).   
  [arXiv:gr-qc/9304007].  

\bibitem{Kodama:2003kk}
  H.~Kodama and A.~Ishibashi,
  Prog.\ Theor.\ Phys.\  {\bf 111}, 29 (2004)
  [arXiv:hep-th/0308128].




  

\end{thebibliography}
\end{document}